%% file: book.tex
\RequirePackage{etoolbox}
\csdef{input@path}%
{%
 {sty/},
 {img/},
}
\documentclass{elsevierbook}

\usepackage{natbib}
\usepackage{amssymb}
\usepackage{amsfonts}
\usepackage{mathtools}
\usepackage{mathrsfs}
\usepackage{graphicx}
\usepackage{hyperref}
\usepackage{booktabs}
\usepackage{cancel}
\usepackage{graphicx}
\usepackage{array}
\newcolumntype{P}[1]{>{\centering\arraybackslash}p{#1}}

\usepackage{xhfill}
\usepackage{ulem}

\newcommand{\bsym}{\boldsymbol}

\DeclareMathOperator*{\minimize}{minimize }

\DeclareMathOperator*{\argmin}{arg\,min}

\begin{document}


  \setcounter{chapter}{6}
  \include{chapter1}


\Backmatter
%
%
%
%
%
%
\bibliographystyle{elsarticle-num}
\bibliography{references}  
%
%
%
%
%
%
%
%
\end{document}

%% file: chapter1.tex
\begin{frontmatter}

\chapter{Hypercomplex Phase Retrieval}\label{chap7}
\subchapter{Kumar Vijay Mishra$^1$, Henry Arguello$^2$, and \\Brian M. Sadler$^3$\\ \small{$^1$United States DEVCOM Army Research Laboratory, Adelphi, MD 20783 USA\\
$^2$Universidad Industrial de Santander, Bucaramanga, Santander 680002 Colombia\\
$^3$The University of Texas, Austin, TX 78712 USA}}

\minitoc
\begin{abstract}
Hypercomplex signal processing (HSP) offers powerful tools for analyzing and processing multidimensional signals by explicitly exploiting inter-dimensional correlations through Clifford algebra. In recent years, hypercomplex formulations of the phase retrieval (PR) problem, wheren a complex-valued signal is recovered from intensity-only measurements, have attracted growing interest. Hypercomplex phase retrieval (HPR) naturally arises in a range of optical imaging and computational sensing applications, where signals are often modeled using quaternion- or octonion-valued representations. Similar to classical PR, HPR problems may involve measurements obtained via complex, hypercomplex, Fourier, or other structured sensing operators. These formulations open new avenues for the development of advanced HSP-based algorithms and theoretical frameworks. This chapter surveys emerging methodologies and applications of HPR, with particular emphasis on optical imaging systems.
\end{abstract}

\begin{keywords}
\kwd{Hypercomplex signal processing} 
\kwd{octonions}
\kwd{phase retrieval}
\kwd{quaternions}
\kwd{sedenions}
\end{keywords}

\end{frontmatter}

\section{Introduction}
In many engineering applications, it is advantageous to model signals of interest using hypercomplex numbers, i.e., elements of algebras defined over the field of real numbers. Unlike vector spaces, which support only addition and scalar multiplication, algebras also permit multiplication between their elements. This additional structure makes hypercomplex representations particularly well suited for multidimensional signal and image processing, as they leverage Clifford algebra to capture and exploit intrinsic correlations across different signal dimensions.

Hypercomplex signal processing (HSP) has found applications in a wide range of areas, including optical imaging, array processing, wireless communications, filtering, and neural networks; see, for example, \cite{miron2023quaternions, buvarp2023quaternion} and the references therein. More recently, HSP has been explored as a framework for addressing the long-standing phase retrieval (PR) problem in high-dimensional settings, especially those encountered in optical imaging. In this chapter, we introduce the fundamental concepts of hypercomplex phase retrieval (HPR) and illustrate representative problem formulations arising in optical imaging applications \cite{jacome2024invitation}.

\subsection{Historical notes}
The origins of hypercomplex numbers date back to the nineteenth century, beginning with the pioneering work of Sir William Rowan Hamilton on quaternion algebra. Hamilton introduced quaternions in a seminal 18-part series published in volumes XXV–XXXVI of \textit{The Philosophical Magazine} between 1844 and 1850 (see, e.g., \cite{hamilton1844ii} and the subsequent papers in the series). This body of work also included the formulation of biquaternions. Shortly thereafter, John T. Graves, inspired by Hamilton’s ideas, discovered the algebra of octonions in 1843. However, Graves published his results only after Arthur Cayley’s influential work on quaternions and their connection to elliptic functions, in which Cayley also described the octonions \cite{cayley1863jacobi}.

These developments were later unified by Leonard E. Dickson, who formalized an iterative algebraic framework now known as the Cayley–Dickson construction. Starting from an algebra endowed with addition, multiplication, conjugation, and vector multiplication, this construction generates a new algebra of twice the dimension by forming the Cartesian product of the algebra with itself. Applied over the field of real numbers, the Cayley–Dickson process produces a sequence of algebras of dimension $2^n$, yielding the complex numbers ($n=1$), quaternions ($n=2$), octonions ($n=3$), sedenions ($n=4$), and higher-order systems. 

In the physical sciences, hypercomplex numbers have attracted sustained interest as mathematical tools for modeling intrinsic properties of space–time, with applications ranging from dynamic vector representations to formulations of the Dirac equation.

\subsection{Desiderata for hypercomplex numbers}
Starting from the two-dimensional real algebras complex numbers, consider $x = a + b\mathrm{i}\in \mathbb{C}  $ where $\mathrm{i}^2 = -1 $ is the imaginary part and $a,b \in \mathbb{R}$. In the two-dimensional algebras, there are also \textit{dual numbers} in which a nilpotent element is used in the basis as $x = a + b\epsilon$ where $\epsilon^2= 0$ and \textit{split-complex } where a hyperbolic imaginary unit is used leading $x = a + b\mathrm{j}$ where $\mathrm{j}^2 = 1$. This algebra satisfies the commutative, associative, and alternative properties. Following the Cayley-Dickson construction, the quaternion numbers are defined as $x = a + b\mathrm{i}+ c\mathrm{j}+ d\mathrm{k} \in \mathbb{H}$  where $a,b,c,d \in \mathbb{R}$ and $\mathrm{i}^2=\mathrm{j}^2=\mathrm{k}^2 = -1$. This algebra is no longer commutative but achieves associative and alternative properties. In this family of four-dimensional algebra, we also have split-quaternions in which $\mathrm{i}^2=-1,\mathrm{j}^2= {\mathrm{k}}^2=1$   which contains zero divisors, nilpotent elements, and nontrivial idempotents. Also, if coefficients $a,b,c,d \in \mathbb{C}$ i.e., $a=a_r + a_i\mathrm{I}, b=b_r + b_i\mathrm{I}, c=c_r + c_i\mathrm{I}, d=d_r + d_i\mathrm{I}$ where $\mathrm{I}^2=-1$  allows commutative property. 

Next, an octonion number $x$ is defined as  {$x=a_0+\sum_{i=1}^{7}a_i \mathrm{e}_i$}, where $a_i$ are real-valued coefficients and $\mathrm{e}_i$ are the \textit{octonion units} such that $\mathrm{e}_i^2=-1$ for $i=1,\dots,7$. Octonion algebra is non-associative and non-commutative. A sedenion  number $x$ is defined as  {$x=a_0+\sum_{i=1}^{15}a_i \mathrm{e}_i$}, where $a_i$ are real-valued coefficients and $\mathrm{e}_i$, $\mathrm{e}_i^2=-1$ for $i=1,\dots,15$, are the \textit{sedenionn units} which form a basis of the vector space of sedenions. Like the octonions, sedenion algebra is non-associative and non-commutative. However, unlike octonions, it is also non-alternative. Further, higher-order hypercomplex numbers, such as trigintaduonions for $n=5$, have also been studied \cite{cariow2014algorithm}.

Table~\ref{tbl:props} compares the algebraic properties of selected hypercomplex number systems. Here, division algebra implies if equation $ax=b$ always has a unique solution (no zero divisors). The norm definite property checks if the norm $xx^*$ is only zero if $x=0$ (positive-definite), or can be zero for $x \neq 0$ (indefinite/isotropic). The idempotent property implies $x^2=x$. It follows that commutativity is lost at $\mathbb{H}$, associativity at $\mathbb{O}$, and both alternativity and division properties are $\mathbb{S}$. From $n=16$ onward, zero divisors are available. For split algebras, $i^2=-1$ is replaced with $j^2=+1$, where $j\neq \pm1$, in the construction resulting in non-division algebras containing non-zero elements whose norm is zero (zero divisors). Biquarternions are similar to split-quarternions, except that they have a complex-valued multiplicative norm.

\begin{table}[t]
\caption{Structural properties of real and split Cayley–Dickson algebras}
\label{tbl:props}
\tiny
\begin{tabular}{|p{1.6cm}|p{0.1cm}|p{1.0cm}|p{0.8cm}|p{0.8cm}|p{0.8cm}|p{1.2cm}|p{1.2cm}|p{1.1cm}|}
\hline \textbf{Number System} & \textbf{$n$} & \textbf{Commutative $xy=yx$} & \textbf{Associative $(xy)z=x(yz)$} & \textbf{Alternative $(yx)x=y(xx)$} & \textbf{Division Algebra}  & \textbf{Zero Divisors} & \textbf{Norm Definite} & \textbf{Contains Idempotents ($\neq 0,1$)}\\
\hline Real ($\mathbb{R}$) & 1 & \checkmark & \checkmark & \checkmark & \checkmark & $\times$ & \checkmark (Positive) & $\times$\\
\hline Complex ($\mathbb{C}$) & 2 & \checkmark & \checkmark & \checkmark & \checkmark & $\times$ & \checkmark (Positive) & $\times$\\
\hline Split-complex & 2 & \checkmark & \checkmark & \checkmark & $\times$ & \checkmark & $\times$ (Indefinite) & \checkmark\\
\hline Quaternions ($\mathbb{H}$) & 4 & $\times$ & \checkmark & \checkmark & \checkmark & $\times$ & \checkmark (Positive) & $\times$\\
\hline Split-quaternions & 4 & $\times$ & \checkmark & \checkmark & $\times$ & \checkmark & $\times$ (Indefinite) & \checkmark\\
\hline Biquaternions & 8 & $\times$ & \checkmark & \checkmark & $\times$ & \checkmark & $\times$ (Complex) & \checkmark\\
\hline Split-biquaternions & 8 & $\times$ & \checkmark & \checkmark & $\times$ & \checkmark & $\times$ (Indefinite) & \checkmark\\
\hline Octonions ($\mathbb{O}$) & 8 & $\times$ & $\times$ & \checkmark & \checkmark & $\times$ & \checkmark (Positive) & $\times$\\
\hline Sedenions ($\mathbb{S}$) & 16 & $\times$ & $\times$ & $\times$ & $\times$ & \checkmark & $\times$ (Isotropic) & \checkmark\\
\hline Trigintaduonions ($\mathbb{T}$) & 32 & $\times$ & $\times$ & $\times$ & $\times$ & \checkmark & $\times$ (Isotropic) & \checkmark\\
\hline
\end{tabular}
\end{table}\normalsize

\subsection{Hypercomplex signal processing}
In signal processing, developing effective modeling, analysis, and processing tools for high-dimensional signals remains a persistent challenge. Over the past three decades, hypercomplex signal processing (HSP) has emerged as a powerful framework for addressing this challenge by leveraging algebraic structures that enable the joint, holistic processing of all signal dimensions, rather than treating each dimension independently as in conventional real- or complex-valued approaches.

One of the most prominent application areas of HSP is image processing, particularly for color imagery. In this context, color images are naturally represented using quaternions, where the red–green–blue (RGB) pixel values at each spatial location are encoded as a single quaternion-valued entity. This representation facilitates unified processing of color information and has been shown to improve performance in a variety of tasks, including color-sensitive filtering via hypercomplex convolution, correlation-based techniques, color edge detection, and principal component analysis (PCA) \cite{miron2023quaternions, chen2019low}.

Beyond quaternions, biquaternions—a modified quaternion algebra that admits commutative multiplication—have been employed in applications such as image painting and edge detection \cite{pei2004commutative}. Using the Cayley–Dickson construction, quaternions can be further extended to eight-dimensional octonions, which have demonstrated promise in multispectral imaging (MSI), where images may comprise seven or more spectral channels. In such settings, each spectral band can be mapped to a distinct imaginary component of an octonion. Octonion-based representations have been successfully applied to MSI denoising through dictionary learning, as well as to image watermarking and color stereo image reconstruction, where two RGB images are stacked to form a six-channel representation encoded within an octonion-valued signal \cite{lazendic2018octonion, yamni2021novel}.

Analogous to real- and complex-valued signal processing, HSP supports spectral representations through hypercomplex Fourier transforms. In particular, quaternion and octonion Fourier transforms (QFT/OFT) map hypercomplex-valued signals from the spatial or temporal domain into corresponding frequency domains \cite{ell2014quaternion}. These transforms underpin a wide range of applications, including color image filtering and motion estimation. More recently, a rich suite of additional HSP tools has been developed, encompassing the short-time Fourier transform (STFT), fractional Fourier transform (FrFT), wavelet transform (WT), linear canonical transform (LCT), bispectral analysis, and deep learning methods. Such developments are especially relevant for phase retrieval (PR) sensing modalities, where the inherent ill-posedness of the classical PR problem is often mitigated through the use of multiple or structured measurements (e.g., Fourier, STFT, FrFT, or wavelet-based measurements) \cite{pinilla2021wavemax, pinilla2024phase, pinilla2025wavemax}.
 
\section{Conventional Complex-Valued Phase Retrieval}
Conventional PR has been significantly researched signal processing problem, where we want to recover a complex-valued signal $\bsym x \in \mathbb{C}^{n}$  given its amplitude-only measurements $\mathbf y\in \mathbb{C}^{m}$ as $\bsym y = \vert\mathbf{A x }\vert^2 $, where the known measurement matrix $\mathbf A \in \mathbb{C}^{m\times n}$.  This problem arises in several areas such as diffractive imaging \cite{pinilla2023unfolding}, X-ray crystallography \cite{pinilla2018phase}, astronomy \cite{fienup1987phase},
and radar waveform design \cite{pinilla2021banraw}. According to the physical application of the PR formulation, the sensing matrix $\mathbf{A}$ is defined accordingly. For instance, in near-field optical imaging \cite{pinilla2023unfolding}, the sensing matrix is the Fourier matrix, and a diffractive optical element (DOE) is employed to reduce the ambiguity of the phaseless Fourier measurements. Usually, to improve the recovery performance, multiple snapshots are acquired varying the DOE. In this case, the acquisition model becomes $\mathbf{y} = \vert\left[\mathbf{D}_1^H\mathbf{F}^H,\dots,\mathbf{D}_L^H\mathbf{F} ^H\right]^H\mathbf{x}\vert^2$ where $\mathbf{F}\in \mathbb{C}^{n\times n}$  is the discrete Fourier transform (DFT) and $\mathbf{D}_\ell \in \mathbb{C}^{n\times n}$ for $\ell = 1,\dots, L$ are diagonal matrices whose entries are the diagonalization of the DOE coding elements and $L$ is the number of acquired snapshots.  

In other applications, for instance, in measuring ultrashort laser pulses \cite{trebino1993using}, the phaseless measurements are obtained via short-time Fourier transform (STFT), where a window $\mathbf{w}$ with compact frequency support of size $R$, i.e., $\Vert \mathbf{w}\Vert_\infty = R$.  Then, the measurements are $\mathbf{y} = \vert \left[\mathbf{W}_1 ^H\mathbf{F}^H,\dots,\mathbf{W}_L^H \mathbf{F}^H\right]^H\mathbf{x}\vert^2$ where $\mathbf{W}_\ell$  is a diagonal matrix with entries given by the $L-$circular shifting of the window $\mathbf{w}$. Thus, multiple sliding window measurements reduce the ill-posedness of the PR problem. Beyond imaging applications, other formulations have been employed, for instance, in audio processing, the problem of recovering phase information after computing the modulus of the wavelet transform (scalogram) is of wide interest \cite{waldspurger2017phase}. Here, the sensing becomes $\mathbf{y} = \vert\left[\bsym{\Phi}_1^H, \dots \bsym\Phi_L^H\right]^H\mathbf{x}\vert^2 $, where $\bsym\Phi_\ell$ is a convolution matrix built upon  {dilated} version of the mother wavelet $\bsym{\phi}_\ell$. 

The PR recovery generally aims to solve the following optimization problem:
\begin{equation}
    \minimize_{\widetilde{\mathbf{x}}\in\mathbb{C}^n} \frac{1}{2m} \sum_{\ell=1}^{m}\mathcal{L}(y_\ell,\vert \langle \widetilde{\mathbf{x}},\mathbf{a}_\ell\rangle \vert^2)\label{eq:conv_pr}
\end{equation}
where $\mathbf{a}_\ell$ is the $\ell$-th row of the sensing matrix and $\mathcal{L}$ is a fidelity measure. Due to the PR model, solving \eqref{eq:conv_pr} includes a factor ambiguity of phase factor $e^{\mathrm{i}\theta}, \theta \in [0,2\pi]$, since $\vert \mathbf{x}\vert =\vert \mathbf{x}e^{\mathrm{i}\theta}\vert$. Thus, the recovery distance is defined as $d(\widetilde{\mathbf{x}},\mathbf{x}) = \min_{\theta}\Vert \mathbf{x}e^{\mathrm{i}\theta}-\widetilde{\mathbf{x}}\Vert_2$. An $\epsilon$-feasible solution of \eqref{eq:conv_pr} is defined by the region $E_\epsilon(\widetilde{\mathbf{x}}) = \{\widetilde{\mathbf{x}}\in \mathbb{C}^n:d(\widetilde{\mathbf{x}},\mathbf{x})<\epsilon\}$.

Considerable efforts have been devoted to the algorithm development (see, e.g., {\cite{vaswani2020nonconvex,pinilla2023unfolding,dong2023phase}} for recent surveys, and references therein). For example, convex optimization-based with extensive recovery theoretical guarantees {\cite{candes2013phaselift}} by converting the quadratic problem to a recovery of a rank-1 matrix via semidefinite programming. While high performance is achieved with these methods, the computational cost becomes infeasible at large-scale signal dimensions. On the other hand, non-convex optimization alleviates the computational complexity by performing gradient descent-like iterations. In these approaches, the selection of the fidelity function $\mathcal{L}$ plays a crucial role. For instance, the seminal paper on the {Wirtinger} flow (WF) algorithm \cite{candes2015phase_2} {employs a quadratic loss} of the squared magnitude measurements. {In \cite{chen2015solving}, the WF convergence is improved by changing the quadratic cost function for a Poisson-based data fidelity function}. Further improvements include using the least square function \cite{zhang2017nonconvex}, or median-truncated functions to obtain a robust algorithm to outliers \cite{zhang2018median}. 
 Another key component of these methods is the initialization algorithm, which is usually performed via spectral methods to compute the leading eigenvector of the measurement matrix.  Based on the vast PR literature, in the following section, we formulate HPR models and algorithms.

\section{Hypercomplex Phase retrieval}
Table \ref{tab:definitions} lists useful definitions for quaternion and octonion signals for which we now introduce PR formulations.
\begin{table}
    \centering
    \caption{Quaternion, octonion, and sedenion identities}
    \begin{tabular}{p{1.4cm}P{3.0cm}P{2.5cm}P{3.0cm}}
    \hline\noalign{\smallskip}
        Definition & Quaternion & Octonion & Sedenion \\
        \noalign{\smallskip}
        \hline
        \noalign{\smallskip}
        & $x\in \mathbb{H},\mathbf{x}\in \mathbb{H}^n$ & $x\in \mathbb{O},\mathbf{x}\in \mathbb{O}^n$ &  $x\in \mathbb{S},\mathbf{x}\in \mathbb{S}^n$ \\
        Conjugation & $x^* = a - b\mathrm{i}-c\mathrm{j}-d\mathrm{k}$ & $x^* =x_0  - \sum_{1}^{7}x_i\mathrm{e}_i$ & $x^* =x_0  - \sum_{1}^{15}x_i\mathrm{e}_i$\\
        Modulus & $\vert {x} \vert = \sqrt{a^2+b^2+c^2+d^2}$  & $\vert {x} \vert = \sqrt{\sum_{i=0}^7 x_i^2}$ & $\vert {x} \vert = \sqrt{\sum_{i=0}^15 x_i^2}$\\
        Inverse ($\vert x \vert^{2} \neq 0$) & $x^{-1} = \frac{x^*}{\vert x \vert^{2}}$ & $x^{-1} = \frac{x^*}{\vert x \vert^{2}}$ & $x^{-1} = \frac{x^*}{\vert x \vert^{2}}$ \\
       Vector Norm  & $\Vert \mathbf{x} \Vert = \sum_{i=0}^{n-1}\vert \mathbf{x}_i\vert$& $\Vert \mathbf{x} \Vert = \sum_{i=0}^{n-1}\vert \mathbf{x}_i\vert$ & $\Vert \mathbf{x} \Vert = \sum_{i=0}^{n-1}\vert \mathbf{x}_i\vert$\\
       \noalign{\smallskip}\hline\noalign{\smallskip}
    \end{tabular}
    \label{tab:definitions}
\end{table}

\subsection{Quaternion PR (QPR): Real-valued sensing matrix}
Consider estimating the quaternion signal $\mathbf{x}\in \mathbb{H}^n$ from quadratic intensity-only measurements $\mathbf{y} = \vert \mathbf{Ax}\vert^2$ where $\mathbf{A}\in\mathbb{R}^{m\times n}$ is the sensing matrix. This problem was first tackled in \cite{chen2022phase_2} and provides several characterizations to determine complex/vector-valued functions in a linear space of finite dimensions, up to a trivial ambiguity, from the magnitudes of their linear measurements. The vector-valued formulation was extended to {quaternion-valued functions}. Its analysis was developed also for affine PR of vector-valued functions, which is a formulation that is encountered in holography \cite{liebling2003local}. However, \cite{chen2022phase_2} did not propose a recovery algorithm.
However, the sensing matrix in this problem is limited to only real values, where the model fails to harness the quaternion multiplication. Instead, it considers $\mathbb{H}$ as $\mathbb{R}^4$ space, thereby reducing QPR to a special case of PR of a real vector-valued signal. Moreover, $\mathbf{A} \in \mathbb{R}^{m\times n}$ leads to more trivial ambiguities that may limit applications of QPR.

\subsection{Quaternion PR: Quaternion-valued sensing matrix} \label{sec:qwf}
In general, HPR entails the estimation of a hypercomplex number measured through a hypercomplex sensing matrix. In this context, unlike \cite{chen2022phase_2}, \cite{chen2023phase} considered a quaternion-valued sensing matrix $\mathbf{A}\in \mathbb{H}^{m\times n}$ to solve the following optimization problem 
\begin{equation}    \minimize_{\widetilde{\mathbf{x}}\in\mathbb{H}^n}f(\mathbf{\widetilde{x}}) =  \frac{1}{2m} \sum_{\ell=1}^{m}(\vert \mathbf{a}_\ell^*\mathbf{\widetilde{x}}\vert^2-y_\ell)^2.\label{eq:qpr_opt}
\end{equation}

 \textbf{$\mathbb{HR}$ Calculus:} The QPR is based on the quaternion vector/matrix derivatives defined below. \vspace{3mm}

  \textit{Definition 1 (Quaternion Rotation) \cite{xu2015theory}}: is defined for any two quartenions $x$ and $\mu$, the transformation $$x^\mu \coloneqq  \mu x\mu ^{-1}$$
  is a 3-dimensional rotation of the vector part of $x$ about the vector part of $\mu$.
\vspace{3mm}

\textit{Definition 2  {(Generalized $\mathbb{HR}$ (GHR) derivatives)}\cite{xu2015theory}}: is defined for a  quaternion $x = a+b\mathrm{i}+c\mathrm{j}+d{\mathrm{k}}$, and a function $f$ with respect to the transformed quaternion $x^\mu$ as 
$$\frac{\partial_l f}{\partial x^{\mu}} = \frac{1}{4}\left(\frac{\partial f}{\partial a}-\frac{\partial f}{\partial b}\mathrm{i}^{\mu}-\frac{\partial f}{\partial c}\mathrm{j}^{\mu}-\frac{\partial f}{\partial c}\mathrm{k}^{\mu}\right).$$
(Without loss in generality, the $\mathbb{HR}$ calculus can be derived from either left or right GHR derivatives.)
\vspace{3mm}

\textit{Definition 3 (Quaternion gradient) \cite{xu2015theory}}: is defined for a scalar function $f(\mathbf{x})$, where $\mathbf{x}\in\mathbb{H}^{n}$, the gradient of $f$ with respect to $\mathbf{x}$ is defined as
$$\nabla_\mathbf{x} f = \left(\frac{\partial f}{\partial \mathbf{x}^\mu}\right)^T,$$ 
where $\frac{\partial f}{\partial \mathbf{x}^\mu} = \left[\frac{\partial f}{\partial \mathbf{x}_1^{\mu}}, \dots, \frac{\partial f}{\partial \mathbf{x}_n^{\mu}}\right]$.

\textbf{Gradient of QPR cost function} To tackle \eqref{eq:qpr_opt}, \cite{chen2023phase} proposed quaternion  WF (QWF) that comprises an initialization stage followed by the recovery. In conventional PR, WF performs gradient-descent-like iterations based on the complex-valued calculus denoted by $\mathbb{CR}$  (also known as the Wirtinger calculus). In QPR, to compute QWF iterations of the cost function in \eqref{eq:qpr_opt}, GHR derivatives are applied as
\begin{align}
    \frac{\partial f(\widetilde{\mathbf{x}})}{\partial\widetilde{\mathbf{x}}}& = \frac{1}{2m} \sum_{\ell=1}^{m} \frac{\partial}{\partial\widetilde{\mathbf{x}}} \left(\vert\mathbf{a}_\ell^*\widetilde{\mathbf{x}}\vert^4- 2y_\ell\vert\mathbf{a}_\ell^*\widetilde{\mathbf{x}}\vert^4 \right)\nonumber\\&=\frac{1}{m} \sum_{\ell = 1}^{m}(\vert \mathbf{a}^*_\ell \widetilde{\mathbf{x}}\vert^2-y_\ell)  \widetilde{\mathbf{x}}^*\mathbf{a}_\ell\mathbf{a}_\ell^*
\end{align}

The initialization estimates $\widetilde{\mathbf{x}}_0$, which has an approximated direction of $\mathbf{x}$.  Then, a spectral method yields the leading eigenvector $\mathbf{v}$ of the Hermitian matrix $\mathbf{Y} = \frac{1}{m}\sum_{\ell=1}^{m}y_\ell \mathbf{a}_\ell \mathbf{a}_\ell^* \in\mathbb{H}^{n\times n}$; note that $\mathbb{E}[\mathbf{Y}] = \mathbf{I}_n + \frac{1}{2}\mathbf{xx}^*$ {where the expectation is  {with} respect to the random rows of the sensing matrix}. Since $\mathbb{E}[y_\ell] = \Vert\mathbf{x}\Vert_2^2$, the \textit{scaling factor} $r = (\frac{1}{m}\sum_{\ell=1}^{m} y_\ell^2)^{1/2}$ is approximately the magnitude of $\mathbf{x}$ when $m$ is sufficiently large. The initial value becomes $\mathbf{\widetilde{x}}^0 = r\mathbf{v}$. To derive the QWF iteration, we rely on the quaternion calculus $\mathbb{HR}$ to perform the derivatives of the cost function \eqref{eq:qpr_opt}.

Then, QWF employs GHR  to compute gradient-descent-like iterations:
\begin{equation}
    \widetilde{\mathbf{x}}^{i+1} =\widetilde{\mathbf{x}}^{i} - \eta \nabla_{\widetilde{\mathbf{x}}}f({\widetilde{\mathbf{x}}}^i),
\end{equation}
for $i=1,\dots,I$ iterations. This algorithm has been proven to converge at linear rates under the distance, \begin{equation}
d(\mathbf{x},\widetilde{\mathbf{x}}) = \min_{ {w}} \Vert\widetilde{\mathbf{x}}-\mathbf{x}w\Vert_2 = \Vert\widetilde{\mathbf{x}}-\mathbf{x}\operatorname{sign}(\mathbf{x}^*\widetilde{\mathbf{x}})\Vert_2,
\end{equation} where $\vert{w}\vert = 1$ is a constant phase factor and $\operatorname{sign}(x) = \frac{x}{\vert x \vert}$, when the quaternion components of the signal are independent~\cite{chen2023phase}. 

\begin{figure}
			\includegraphics[width=1.0\linewidth]{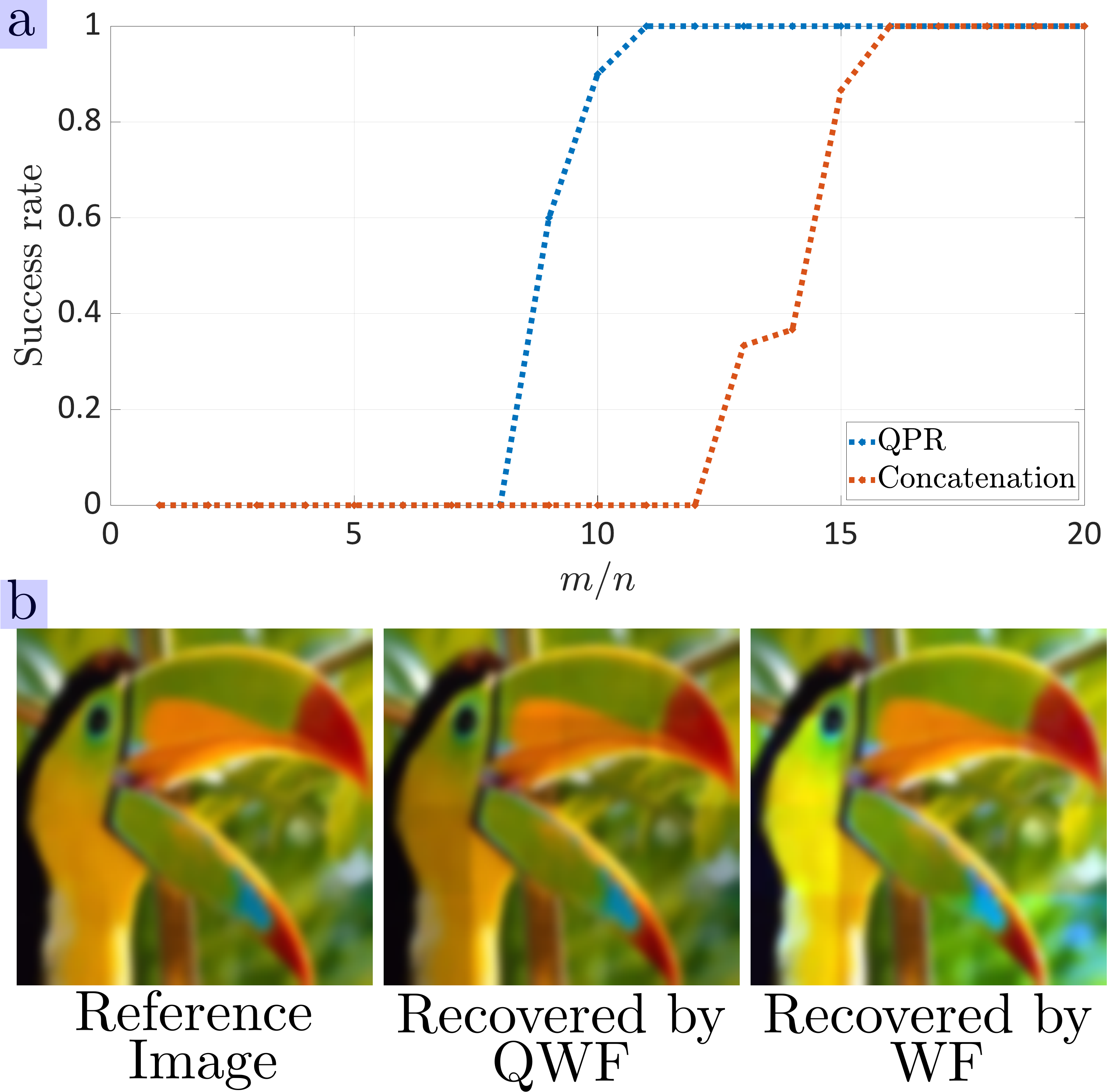}
				\caption{(a) Recovery performance of QWF for different sample complexity and WF algorithm by concatenating the four dimensions of the quaternion signal. {For the quaternion setting, the sensing matrix was drawn from a quaternion normal distribution. For the concatenation, the same was drawn from a complex normal distribution}. The experiments are performed for $n=500$. (b) Illustration of QPR recovery using a red-green-blue (RGB) image. Here, we set each color channel as the imaginary parts of a quaternion signal and reconstruct it with QWF. For comparison, by concatenating each color channel and applying WF, the recovery is poorer than QPR. In both cases, reconstruction was performed in patches of $N=32$ such that $n=1024$ and $m/n = 15$.} 
                \label{fig:transition_qpr}       
			\end{figure}

Figure~\ref{fig:transition_qpr}a plots the performance of QWF by employing $\mathbb{HR}$ calculus. where the sample complexity $m/n$ was varied from 100 Monte-Carlo simulations. The reconstructions with real RGB image data are shown in Figure~\ref{fig:transition_qpr}b. Results suggest that QWF allows a better performance for both synthetic signals and real RGB signals.

Other WF-like variations of QPR also exist. For instance, quaternion truncated WF (QTWF) changes the cost function in \eqref{eq:qpr_opt} to $f(\widetilde{\mathbf{x}}) = \frac{1}{m}\sum_{\ell=1}^{m}y_\ell\log(\vert\mathbf{a}_\ell^*\widetilde{\mathbf{x}}\vert)-\vert\mathbf{a}_\ell^*\widetilde{\mathbf{x}}\vert $ and performs GHR to develop the gradient descent iterations. In this case, the gradient is trimmed to be more selective by truncation. Similarly, the QWF variants for base truncated amplitude WF \cite{pinilla2023unfolding} 
and pure quaternion signal has also been defined \cite{chen2023phase}. 

A different QPR problem in \cite{shao2014double} considers the quaternion gyrator domain for a double-image encryption application. This transformation is a linear canonical integral transform which is viewed as an extension 
\noindent of Fourier analysis. It admits two quaternion signals $\mathbf{x}_1$ and $\mathbf{x}_2$ such that $\mathbf{y} = \vert \mathbf{A}_1 \mathbf{x}_1\vert^2= \vert \mathbf{A}_2 \mathbf{x}_2\vert^2$  where $\mathbf{A}_1$ and $\mathbf{A}_2$ are built upon \textit{gyrator transform} \cite{shao2013quaternion} with different rotation angles and phase mask that perform the encryption of the images. 

\subsection{Octonion PR (OPR)}
Ocotnions have garnered attention in applications including MSI processing \cite{lazendic2018octonion}, wherein each pixel 7-color channel image has a vector-valued representation such that each channel corresponds to different complex-variable dimensions. Octonions are employed for color-stereo image analysis to represent two 3-color channel images in a different imaginary dimension. The mutual processing along the color channels and two stereo images has been shown to improve the analysis. Recently, there has been broad interest in the recovery of the complex MSI represented using octonion-valued signals from its phaseless measurements. 

Consider the octonion-valued signal $\bsym{x} \in \mathbb{O}^{n}$ and its phaseless measurements  $\mathbf{y} = \vert\mathbf{A}\mathbf{x}\vert^2\in \mathbb{R}_+^m$ where $\mathbf{A}\in\mathbb{O}^{m\times n}$ is the octonion-valued sensing matrix. Note that, given a unit octonion $q$ such that $\vert q\vert = 1$, the signal $\mathbf{x}$ scaled by a global right pure octonion factor leads to the same $\mathbf{y}$, i.e., $\vert\mathbf{Ax}q\vert^2=\vert\mathbf{Ax}\vert^2$. The octonion algebra is non-commutative and, hence, $\vert \mathbf{A}q\mathbf{x}\vert^2\neq\vert\mathbf{Ax}\vert^2$. Then, OPR aims to recover $\mathbf{x}$ up to a \textit{trivial ambiguity} of only on the right octonion phase factor \cite{jacome2024octonion}. 

 \textbf{Octonion real matrix representation}
				Unlike quaternion algebra, a real-matrix representation of an octonion number does not exist. However, \cite{rodman2014hermitian} proposed a pseudo-real matrix representation that has been successfully employed by many octonion-valued signal  {applications. To obtain} To obtain this representation, define the real representation of the octonion number $x\in \mathbb{O}$ as $\aleph(x) = [x_0,x_1,\dots,x_7]^T\in \mathbb{R}^{8}$. Then, the injective mapping $\gimel : \mathbb{O} \rightarrow \mathbb{R}^{8\times 8}$ is the {real matrix representation} of an octonion number \cite{rodman2014hermitian}: 
\par\noindent\small
\begin{equation}
    \gimel(x) =\left[\begin{smallmatrix}
x_0 & -x_1 & -x_2 & -x_3 & -x_4 & -x_5 & -x_6 & -x_7 \\
x_1 & x_0 & x_3 & -x_2 & x_5 & -x_4 & -x_7 & x_6 \\
x_2 & -x_3 & x_0 & x_1 & x_6 & x_7 & -x_4 & -x_5 \\
x_3 & x_2 & -x_1 & x_0 & x_7 & -x_6 & x_5 & -x_4 \\
x_4 & -x_5 & -x_6 & -x_7 & x_0 & x_1 & x_2 & x_3 \\
x_5 & x_4 & -x_7 & x_6 & -x_1 & x_0 & -x_3 & x_2 \\
x_6 & x_7 & x_4 & -x_5 & -x_2 & x_3 & x_0 & -x_1 \\
x_7 & -x_6 & x_5 & x_4 & -x_3 & -x_2 & x_1 & x_0
 \end{smallmatrix}\right].\nonumber
\end{equation}\normalsize
Both representations $\aleph$ and $\gimel$ are also easily extended to vector/matrix octonion variables {i.e.,  consider $\mathbf{A}\in\mathbb{O}^{m\times n}$, we have that $\aleph(\mathbf{A}) \in \mathbb{R}^{8n\times m}$ and $\gimel({\mathbf{A}})\in\mathbb{R}^{8m\times 8n}$. Consider $\mathbf{x} \in \mathbb{O}^{n}$ and $\mathbf{A} \in \mathbb{O}^{m\times n}$, it holds $\aleph(\mathbf{Ax}) = \gimel(\mathbf{A})\aleph(\mathbf{x})$ and $\Vert\mathbf{x}\Vert = \Vert \aleph(\mathbf{x})\Vert_2$.}

\textbf{OPR distance metric}
To measure the error between the estimated octonion signal and the ground truth signal, define the distance $d(\mathbf{x},\mathbf{x}^\ast) = \min_{z}\Vert \mathbf{x}^{\ast}-\mathbf{x}z\Vert$ where $z \in \{z \in \mathbb{O}\vert \vert z\vert = 1\}$ is only-phase octonion factor. We represent this distance in terms of the pseudo-real-matrix representation of octonions as $d(\mathbf{x}^\ast,\mathbf{x}) =  \min_{z}\Vert\aleph(\mathbf{x}^\ast)- \gimel(\mathbf{x}))\aleph(z)\Vert$, using the property $\Vert \mathbf{x}\Vert = \Vert \aleph(\mathbf{x})\Vert$. After some tedious algebra, we obtain $d(\mathbf{x},\mathbf{x}^{\ast}) = \Vert\aleph(\mathbf{x}^\ast)- \gimel(\mathbf{x})) g(\mathbf{x}^{\ast})\Vert$, where 
\begin{equation}
     g(\mathbf{x}^{\ast}) = \operatorname{sign}\left(\left(\gimel(\mathbf{x})^T\aleph(\mathbf{x}^{\ast})\right)\left(\gimel(\mathbf{x})^T\gimel(\mathbf{x}\right)^{-1}\right).
\end{equation}

As before, the octonion WF (OWF) \cite{jacome2024octonion} employs the spectral initialization, wherein our goal is to obtain the initial estimate of the true signal by computing the leading eigenvector of the octonion-valued matrix $\mathbf{Y} = \frac{1}{m}\sum_{\ell=1}^{m} {{y}_\ell} \mathbf{a}_\ell\mathbf{a}_\ell^* \in \mathbb{O}^{n\times n}$. This may be achieved by solving an octonionic right eigenvalue decomposition. This decomposition has been solved earlier for small octonion-valued matrices ($n<4$) but it cannot be extended to larger matrices. Therefore, \cite{jacome2024octonion} adapted the power method for the right quaternion eigenvalue decomposition to compute the leading eigenvalue of $\mathbf{Y}$. This method relies on the real-matrix representation of octonion numbers. Using power iterations over the real matrix representation, the inverse real representation $\aleph^{-1}(\cdot)$ is employed to yield the equivalent octonion leading eigenvalue. 

The octonion algebra is also non-associative. Hence, it lacks a clear definition of octonion derivatives, including chain rule, high-dimensional gradients, and gradient-based methods such as the WF.  Optimization methods that employ octonion representation (e.g., in singular value decomposition (SVD) or deep octonion neural networks) usually resort to the real-matrix representation to perform optimization over the real-valued variable.

 Inspired by this approach, \cite{jacome2024octonion} employed this representation to solve the following OPR problem:
\begin{align}
\minimize_{\widetilde{\mathbf{x}}\in \mathbb{O}^{n}} \sum_{\ell=1}^{m} {\frac{1}{2}}\left(\vert\mathbf{a}_\ell^* \widetilde{\mathbf{x}}\vert^2- {y}_\ell\right)^2.
\end{align}
Employing the real matrix representation, OPR is formulated as
\begin{align}
    \widetilde{\mathbf{x}} &= \aleph^{-1}\left(\argmin_{{\mathbf{x}}\in \mathbb{R}^{8n}} \sum_{\ell=1}^{m}  {\frac{1}{2}}\left(\Vert\gimel\left(\mathbf{a}_\ell^*\right)\aleph({\mathbf{x}})\Vert_2^2- {y}_\ell\right)^2\right)\nonumber\\&=  \aleph^{-1}\left(\argmin_{{\mathbf{x}}\in \mathbb{R}^{8n} }f({\mathbf{x}}) \right),\label{eq:problem_real}
\end{align}
where the $\ell_2$ norm comes from the observation that the norm of an octonion variable is the norm of its real representation. Then, \eqref{eq:problem_real} is solved by applying the gradient descent steps on the cost function as 
\begin{align}
\nabla_{ {\mathbf{x}}} f( {\mathbf{x}}) = \sum_{\ell=1}^{m}(\Vert\gimel\left(\mathbf{a}_\ell^*\right)\aleph( {\mathbf{x}})\Vert_2^2-\mathbf{y}_\ell)\gimel\left(\mathbf{a}_\ell^*\right)^T\gimel\left(\mathbf{a}_\ell^*\right)\aleph( {\mathbf{x}}).
\end{align}
 Then, the OWF update process in the $i$-th iteration, where $i\in \{1,\dots,I\}$ such that $I$ is the maximum number of iterations, becomes
\begin{equation}
    \widetilde{\mathbf{x}}^{(i)} = \widetilde{\mathbf{x}}^{(i-1)} - \alpha \nabla f(\widetilde{\mathbf{x}}^{(i-1)})
\end{equation}
where $\alpha$ is a suitable selected gradient step size. Finally, we obtain the octonion signal by computing the inverse real representation of $\widetilde{\mathbf{x}}^{(I)}$ as $\widetilde{\mathbf{x}} = \aleph^{-1} (\widetilde{\mathbf{x}}^{(I)})$.

\begin{figure}
			\includegraphics[width=0.95\linewidth]{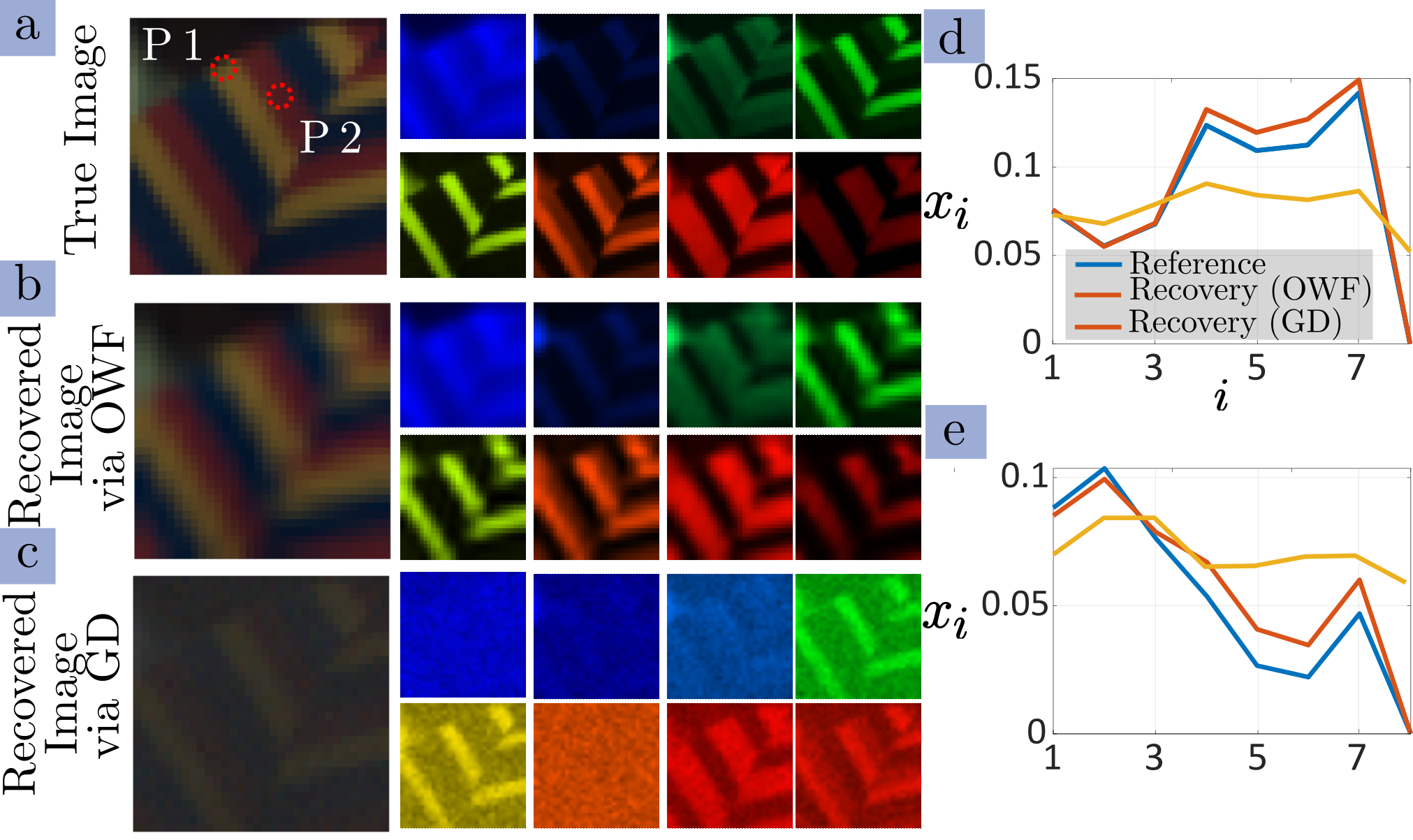}
				\caption{Reconstruction of real data with OWF. (a) RGB representation of the eight-channel spectral image and its individual eight components on the right panel. (b) The OWF-reconstructed image and components. The PSNR of the recovered image is 39.007 [dB]. (c) The GD-based reconstructed image and components.  The PSNR of the recovered image is 24.1615 [dB]. Recovered octonion-valued numbers from two coordinates (c) (15,15) and (d) (10,10), labeled `P1' and `P2', respectively \cite{jacome2024octonion}. } 
                \label{fig:visual}
                \end{figure}

\begin{figure}
                \includegraphics[width=0.95\linewidth]{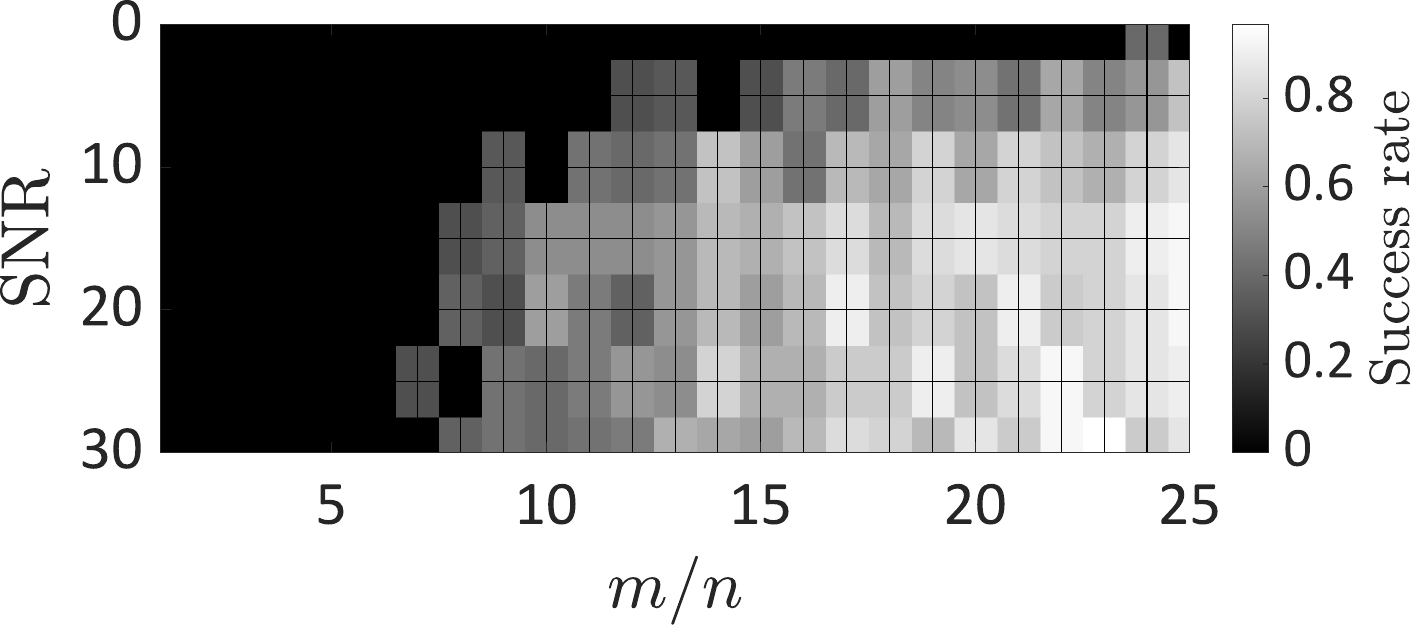}
				\caption{Success rate of OWF for different values of sampling complexity $m/n$ with $n=30$ with measurements under additive Gaussian noise with a signal-to-noise ratio varying from 0 to 30 \cite{jacome2024octonion}.}    \vspace{3mm}
				\label{fig:snr}
			\end{figure}

Figure \ref{fig:visual} presents an application of OWF for MSI. Here, the spectral image (Figure \ref{fig:visual}a) from the CAVE multispectral image dataset was used by cropping it to $32\times 32$ pixels and selecting eight equispaced spectral bands from 31 original spectral bands ranging from 400 to 700 nm.  Figure~\ref{fig:snr} shows OWF recovery for noise measurements for different SNR values. 
                
Each quaternion (octonion) multiplication requires 16 (64) real multiplications \cite{cariow2020algorithm}. The standard WF algorithm has $\mathcal{O} (mn^2\log (1/\epsilon))$ complexity to achieve an accuracy of $d(\mathbf{x},\widetilde{\mathbf{x}}) \leq \epsilon \Vert \mathbf{x} \Vert_2$ \cite{candes2015phase_2}. In comparison, the computational complexities of QWF and OWF are of the order $\mathcal{O} (16^3mn^2\log (1/\epsilon))$ and $\mathcal{O} (64^3mn^2\log (1/\epsilon))$, respectively. We refer the interested reader to \cite{cariow2020fast,cariow2021fast,cariow2023fast} for research on addressing the high computational complexity of quaternion and octonion signal processing techniques.

 \section{Emerging HPR Applications}
Traditional HPR problems employ measurements using a sensing matrix drawn from a random distribution. However, analogous to the conventional PR, measurements may also be taken using other HSP tools such as hypercomplex Fourier or other matrices. We list some of these emerging applications below.
\subsection{Hypercomplex Fourier PR}
\begin{figure}
				\includegraphics[width=0.8\linewidth]{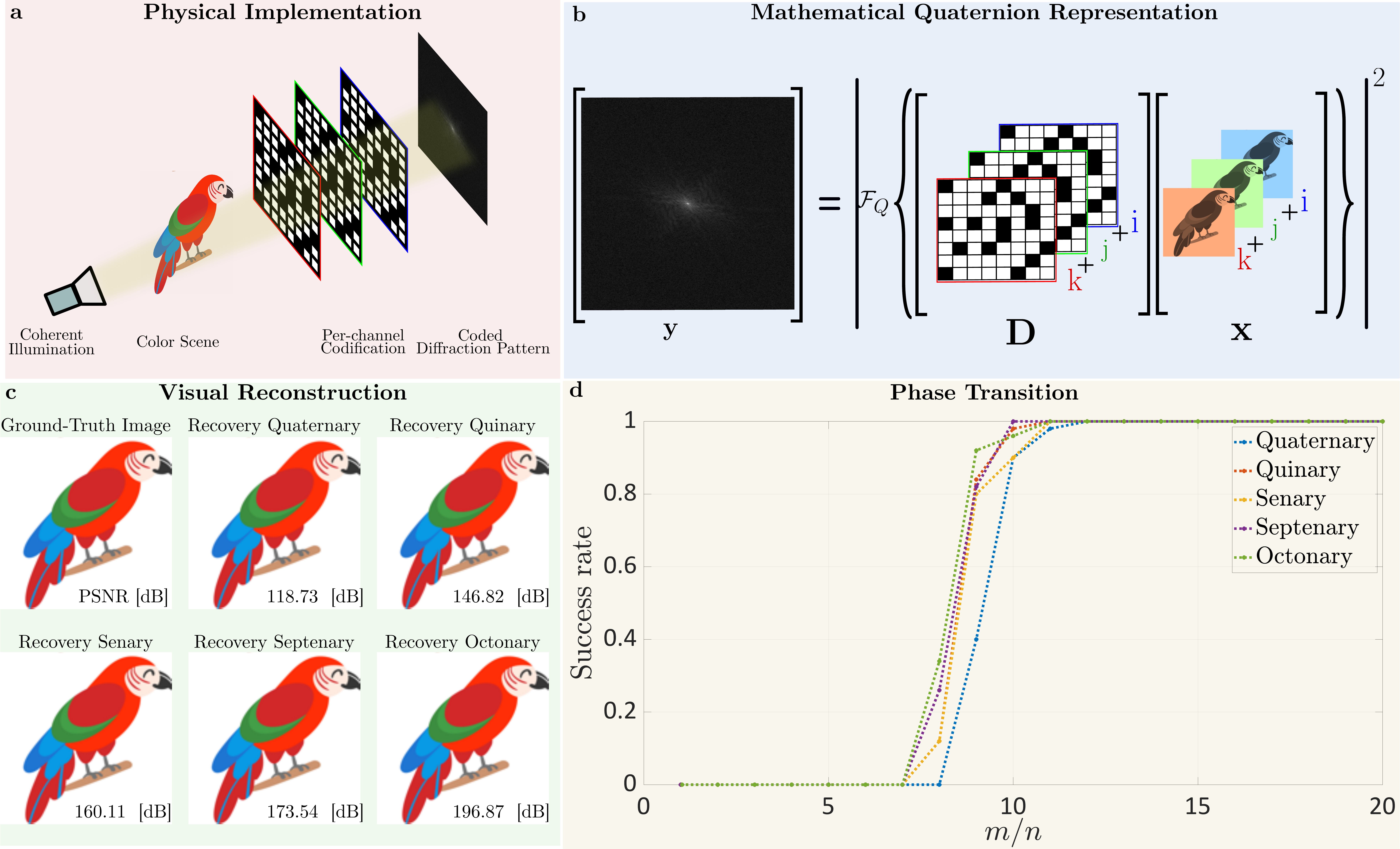}
				\caption{(a) Illustration of the physical measurement set-up of Fourier QPR in optical imaging where a coherent illumination diffracts from an object. A DOE codifies the scene. Considering far-zone propagation, the sensor measures the magnitude of the coded scenes via QDFT. (b) Quaternion modeling of Fourier phase retrieval, where the sensing matrix contains both the per-channel codification of the DOE and the quaternion Fourier transform. (c) Fourier QPR recovery of an RGB image with a different number of coding elements from $d=4$ to $d=8$ using QWF. Here, $m/n=L=10$ and $n=128^2$. (d) Phase transition of QWF for Fourier QPR by varying the sample complexity $m/n$ and different values of $d$. Here, we set $n=1000$.}
				\label{fig:qfpr}
			\end{figure}

Recovering phase information of the light has a plethora of applications including object detection, microscopy, and X-ray crystallography. In diffractive optical imaging (DOI), the sensing process is modeled according to the propagation zone \cite{pinilla2023unfolding}. For instance, far-zone DOI comprises a DOE that modulates the input wavefront, and propagation is modeled by the Fourier transform of the coded wavefront (Figure \ref{fig:qfpr}a). 
While we illustrate only the far-zone case here, other zones (mid- and near-) are addressed similarly using a suitable sensing matrix.

Consider a two-dimensional (2-D) DOE $\mathbf{T}^k \in \mathbb{C}^{N\times N}$, where $k$ is the snapshot index $k=1,\dots,L$. 
Every entry of the DOE is usually a random variable on the set of possible coding elements $\mathcal{D}=\{c_1,\dots,c_d\}$ with probability $\frac{1}{d}$ for each variable and $d$ denotes the number of possible coding elements. In conventional complex-valued PR \cite{candes2015phase_2}, the set of coding variables are $\mathcal{D} = \{1,-1,-\mathrm{i},\mathrm{i}\}$. There are several recent works on the uniqueness guarantees of PR obtained by properly designing the DOE \cite{pinilla2023unfolding}. The sensing matrix here is $\mathbf{A} = \left[(\mathbf{D}^1)^H \mathbf{F}^H,\dots,(\mathbf{D}^L)^H \mathbf{F}^H\right]^H$, where $\mathbf{D}^k \in \mathbb{C}^{n\times n}$ is a diagonal matrix with the vectorized DOE $\mathbf{T}^k$ and $\mathbf{F}\in\mathbb{C}^{n\times n}$  is a 2-D DFT matrix.

 The hypercomplex version of this problem leads to new algorithmic opportunities by harnessing hypercomplex algebras. Thus, the Fourier QPR is formulated using a specific definition of the 2-D QDFT matrix $\mathbf{F}_Q$. {In quaternion signal processing, the Quaternion Fourier Transform (QFT) \cite{ell2014quaternion} does not have a unique definition due to the position of the exponential or the pure quaternion unit employed}. For Fourier QPR, we employ the 2-D QFT (or two-sided QFT). Consider the continuous-valued signal $f(\mathbf{x}): \mathbb{R}^2\rightarrow \mathbb{H}, \mathbf{x} = [x_1,x_2] \in \mathbb{R}^2$, whose QFT $F(\mathbf{u})=\mathcal{F}_Q\{f(\mathbf{x})\}$, where $F(\cdot)$ is a function of the quaternion Fourier domain variable $\mathbf{u} = [u_1,u_2] \in \mathbb{R}^2$. The QFT becomes
\begin{equation}
   F(\mathbf{u}) = \int_{\mathbb{R}^2}e^{-\mathrm{i}2\pi u_1x_1}f(\mathbf{x})e^{-\mathrm{j}2\pi u_2 x_2}d^2\mathbf{x}.
\end{equation}
The corresponding $(N,N)$-point 2-D quaternion DFT (QDFT) of the discrete quaternion signal  {$f(q,b)$} is 
\begin{equation}
    F_Q(r,s) =  {\frac{1}{N^2}} {\sum_{q=0}^{N-1}\sum_{b=0}^{N-1}}e^{-\mathrm{i}2\pi rq/N}f(q,b)e^{-\mathrm{j}2\pi sb/N}.
\end{equation}

Define the 2-D QDFT matrix $\mathbf{F}_Q \in \mathbb{H}^{n\times n}$ for a vectorized quaternion image $\mathbf{x}\in\mathbb{H}^{n}$, $n=N^2$, as $\mathbf{F}_Q = (\widetilde{\mathbf{F}}_Q^{(\mathrm{i})}\otimes  \widetilde{\mathbf{F}}_Q^{(\mathrm{j})} )\mathbf{x}$ where $\otimes$ denotes the Kronecker product. Here, $\widetilde{\mathbf{F}}_Q^{(\mathrm{\mu})}\in\mathbb{H}^{N\times N}$ denotes the N-point 1-D QDFT matrix for the \textit{quaternion unit} $\mathrm{\mu}$, whose entries are $\widetilde{\mathbf{F}}_Q^{(\mathrm{\mu})}(r,s) =\frac{1}{\sqrt{N}}e^{-\mu2\frac{\pi}{N} rs} $ and $\otimes $ denotes the Kronecker product. This matrix formulation of the 2-D QDFT is possible because of the separability of the two-sided QDFT. Computational implementations of the 2-D QDFT via quaternion fast FT (QFFT) are also available.

Denote the set of possible quaternion coding variables by  {$\mathcal{D}_Q = \{q_1,\dots,q_d\}$},  {where $q_i \in \mathbb{H}$ for $i=1,\dots,d$}. Then, the DOE of the $k$-th snapshot $\mathbf{T}^k_Q \in \mathbb{H}^{N\times N}$ is built upon the quaternion coding variables. Assume $\mathbf{D}^k_Q \in \mathbb{H}^{n\times n}$ to be the diagonalization of the quaternion DOE. Based on this formulation, the Fourier QPR measurements are
\begin{equation}
    \mathbf{y} = \left\vert \overbrace{\left[(\mathbf{D}^1_Q)^H\mathbf{F}_Q^H,\dots,(\mathbf{D}^L_Q)^H\mathbf{F}_Q^H\right]^H}^{\mathbf{A}\in \mathbb{H}^{Ln\times n}}\mathbf{x}\right\vert ^2 
\end{equation}
where $\mathbf{y} \in \mathbb{R}^{nL}$ are the phaseless measurements; see Figure \ref{fig:qfpr}b) for a visual representation of this formulation. 

 To recover the quaternion signal $\mathbf{x}$ from $\mathbf{y}$, we solve the optimization problem \begin{equation}
     \minimize_{\widetilde{x}\in\mathbb{H}^n} \frac{1}{2nL} \sum_{\ell=1}^{ {nL}}(\vert\mathbf{a}_\ell^*\widetilde{\mathbf{x}}\vert^2-y_\ell)^2,
 \end{equation} where $\mathbf{a}_\ell = \mathbf{f}_{\alpha_\ell}^H(\mathbf{D}_Q^{k_\ell})^H$ is a row of the QDFT matrix with $\alpha_\ell = \lfloor \frac{\ell}{n}\rfloor+1$ and $k_\ell = ((\ell -1)\mod n) +1$. This formulation is also solved using QWF. 
 
 Figure \ref{fig:qfpr}c employs real data with an RGB image of size $N=128$. Here, $n=128^2=1024$ and $L=10$. 
An increase in the number of coding variables improves the reconstruction quality.  
Figure \ref{fig:qfpr}d shows a phase transition performance of QWF for the Fourier QPR. The results are averaged over $100$ Monte-Carlo simulations for recovering a random quaternion signal with $n=1000$. In Figure \ref{fig:qfpr}c we tested the algorithm with real data, an RGB image. Although all results produce satisfactory reconstructions, increasing the number of coding variables improves the reconstruction quality.  {Figure \ref{fig:qfpr}d shows a phase transition performance of QWF, wherein we declare success when $d(\mathbf{x},\widetilde{\mathbf{x}})<1e-5$.} The results show that increasing the number of coding variables allows for better performance in QWF. This allows us to conclude that using quaternion approaches, increases the coding richness compared with the complex-valued Fourier PR which translates to better recovery performance. 
 
 The Fourier QPR may also be extended to octonions by following the octonion Fourier transform (OFT) \cite{blaszczyk2020discrete} defined for tri-variate functions. This enables novel 3-D PR applications, such as 3-D object reconstruction, structured light 3-D object sensing, and digital holography.  {Similar to QFT, OFT has no unique definition. Here, we follow the definition in \cite{blaszczyk2017octonion} because of its several useful properties (similar to complex-valued FT) such as Hermitian symmetry, shift theorem, Plancherel-Parseval theorem, and derivative theorem.} Consider the octonion continuous-valued signal $f(\mathbf{x}): \mathbb{R}^3\rightarrow \mathbb{H}, \mathbf{x} = [x_1,x_2,x_3] \in \mathbb{R}^3$, whose OFT $F(\mathbf{u})=\mathcal{F}_O\{f(\mathbf{x})\}$, where ${F}(\cdot)$ is a function of the octonion Fourier domain variable $\mathbf{u} = [u_1,u_2,u_3] \in \mathbb{R}^3$. The OFT is given by
\begin{align}
\mathcal{F}_O\{(f(\mathbf{x})\}= F(\mathbf{v}) = \int_{\mathbb{R}^3} f(\mathbf{x})e^{-\mathrm{e}_1 2\pi f_1 x_1}e^{-\mathrm{e}_2 2\pi f_2 x_2}e^{-\mathrm{e}_4 2\pi f_3 x_3}d^3\mathbf{x}.
\end{align}
 
 Here, only the imaginary units $e_1,e_2,e_4$ are used because, following the power expansion of the exponential function, this product yields a full octonion function. Define the octonion DFT (ODFT) for a discrete octonion signal $f(\mathbf{n})=f(n_1,n_2,n_3)$ with  {$n_1,n_2,n_3 \in \{0,\dots, N-1\}$ as
\begin{align}
     F_O(k_1,k_2,k_3) =\frac{1}{N^3}\sum_{n_1=0}^{N-1}\sum_{n_2=0}^{N-1}\sum_{n_3=0}^{N-1}f(n_1,n_2,n_3)e^{-\mathrm{e}_1 2\pi k_1 n_1/N}e^{-\mathrm{e}_2 2\pi k_2 n_2/N}e^{-\mathrm{e}_4 2\pi k_3 n_3/N}.
 \end{align}}
 Similar to Fourier QPR, we have the corresponding ODFT sensing matrix $\mathbf{F}_O \in \mathbb{O} ^{n \times n}$. The DOE matrix $\mathbf{T}_O^k\in\mathbb{O}^{N\times N}$ has more coding variables. The design of optimal DOE to guarantee unique recovery in Fourier HPR is currently an unaddressed problem.

\subsection{Hypercomplex STFT PR}
The STFT of a signal provides the spectral characteristics localized in time. It is obtained by partitioning the signal into overlapping segments and then taking the windowed Fourier transform of each segment. Conventional PR has shown that, for the same number of measurements, the STFT magnitude measurements lead to better PR performance than an over-sampled DFT. The STFT PR applications include ultrashort laser pulses or Fourier ptychography. Traditional complex-valued STFT PR problem considers a discrete-time signal $\mathbf{x} = [x_1,\dots, x_n]^T \in \mathbb{C}^{n}$ and window $\mathbf{w} = [w_1,\dots,w_T]^T\in\mathbb{C}^{T}$ such that the magnitude-only measurements using the discrete-time STFT of the $\mathbf{x}$ respect to the window $\mathbf{w}$ are \begin{equation}
    \mathbf{Y}_{r,s} = \left\vert\sum_{q=1}^{N} \mathbf{x}[q] \mathbf{w}[rL-q]e^{-\mathrm{i}2\pi sq/N}\right\vert^2,\end{equation} $s = 1,\dots,N$, $r = 1,\dots,R$,  where $R = \left \lceil{\frac{N+T-1}{L}
}\right \rceil$ is the number of short-time sections and  {$\mathbf{w}[n]$ is periodically extended over the boundaries of the window}. The choice of the window affects the uniqueness guarantees of recovery. 

In the case of STFT QPR, the continuous-valued quaternion STFT (QSTFT) of a quaternion  {2-D} signal $f(\mathbf{x}) \in  \mathbb{H}$ respect to the window function $w(\mathbf{x}) \in \mathbb{H}$ is 
\begin{align}F_{QSTFT}(\mathbf{v},\mathbf{u})=  \int_{\mathbb{R}^2} f(\mathbf{x}) w(\mathbf{x}-\mathbf{u}) e^{-\mathrm{i}2\pi x_1v_1} e^{-\mathrm{j}2\pi x_2v_2}d^2\mathbf{x},\end{align}
where $\mathbf{v}$ and $\mathbf{u}$ are variables in the STFT domain. Its corresponding discrete version with a quaternion window $w(p,q) \in \mathbb{H}$  and a discrete signal $f(p,q)\in \mathbb{H}$, $p,q \in {1,\dots,N}$, is 
\begin{align}&F(\mathbf{v},\mathbf{u}) \nonumber\\&=  {\frac{1}{N^2}\sum_{q=0}^{N-1}\sum_{b=0}^{N-1}}f(q,b)w(p-u_1,q-u_2)e^{-\mathrm{i}2\pi qr/N}e^{-\mathrm{j}2\pi sb/N}.\end{align}
In a matrix representation, the QSTFT measurements are $\mathbf{Y}_{r,s} =  \mathbf{f}_s^*\mathbf{W}_r\mathbf{x}$ where $\mathbf{W}_r\in \mathbb{H}^{N\times N}$ is a diagonal matrix created by shifting the quaternion vector $\mathbf{w}_r\in \mathbb{H}^T$ across the diagonal and $\mathbf{f}_s^*$ is the $s$-th row of the 2-D QDFT matrix. The QSTFT PR problem entails recovering $\mathbf{x}$ by solving the optimization problem: 
\begin{equation}\minimize_{\widetilde{\mathbf{x}}\in \mathbb{H}^N} \frac{1}{2nL} \sum_{r=1}^{R}\sum_{s=1}^{N}(\vert\mathbf{f}_s^*\mathbf{W}_r\widetilde{\mathbf{x}}\vert^2-\mathbf{Y}_{r,s})^2.\end{equation}

The STFT OPR is similarly defined using the octonion STFT (OSTFT) for 3-D signals as 

\begin{align}
    F_{OSTFT}(\mathbf{v},\mathbf{u})=\int_{\mathbb{R}^3}f(\mathbf{x})w(\mathbf{x}-\mathbf{u}) e^{-\mathrm{e}_1 2\pi x_1v_1} e^{-\mathrm{e}_22\pi x_2v_2}e^{-\mathrm{e}_42\pi x_3v_3}d^3\mathbf{x}.
\end{align}

\subsection{Hypercomplex Wavelet PR}
Wavelet PR has shown promising applications in audio processing, and its hypercomplex variant may be formulated via the quaternion and octonion wavelet transforms. The quaternion wavelet transform (QWT) of a quaternion-valued signal $f(\mathbf{x}):\mathbb{R}^2\rightarrow \mathbb{H}$ with respect to the mother \textit{quaternion wavelet} $\psi(\mathbf{x}):\mathbb{R}^2\rightarrow\mathbb{H}$ for a scaling factor $a$, a rotation angle $\theta$, and the coordinates in the wavelet domain $\mathbf{b} \in \mathbb{R}^2$ is 
\begin{equation}
F_{QWT} (\psi,a,\theta,\mathbf{b}) = \int_{\mathbb{R}^2} f(\mathbf{x})\frac{1}{a}\psi\left( \mathbf{R}_{-\theta} \frac{1}{a} \left(\mathbf{x-b}\right) \right)d\mathbf{x}^2,\end{equation}
where $\mathbf{R}_\theta \in \mathbb{R}^{2\times 2}$ is the rotation matrix on $\mathbb{R}^2$ defined as $\mathbf{R}_\theta = \left(\begin{smallmatrix}
    \cos(\theta) & \sin(\theta)\\-\sin(\theta) & \cos(\theta)
    \end{smallmatrix}\right)$. We can interpret this definition as the convolution of the scaled and rotated mother quaternion wavelet $\psi(\mathbf{x})$ with the 2-D quaternion signal ${f}(\mathbf{x})$. It is instructive to define the family of wavelets $\{\psi^k(\mathbf{x})\}_{k=1}^L$, where $\phi^k(\mathbf{x}) = \frac{1}{a^k}\psi\left( \mathbf{R}_{-\theta^k}\frac{\mathbf{x}}{a^k}\right) $ and $L$ is the number of transformations of the mother wavelet. Then, the QWT becomes $F_{QWT}(\psi,\mathbf{b}) = \{f(\mathbf{x})\star\phi_j(\mathbf{b})\}_{k=1}^{L}$ where $\star$ is the convolution operation. The discrete QWT (DQWT) considers the sampled versions of the signal as of $f(p,q)$ and the mother wavelet transformation $\psi^k(p,q)$, $p,q =1,\dots, N$, as 
    \begin{equation}
         F(r,s) = \sum_{p=1}^{N}\sum_{q=1}^{N} f(p,q)\psi^k(r-p,s-q),
    \end{equation}
    for $r,s=1,\dots,N$. 
    
In the hypercomplex wavelet QPR problem, the magnitude-only measurement vector becomes $\mathbf{y} = \left\vert\left[(\bsym{\psi}^1\star\mathbf{f})^T,\dots,(\bsym{\psi}^L\star\mathbf{f})^T\right]^T \right\vert^2$, where $\mathbf{f}\in\mathbb{H}^n$ is the vectorization of the quaternion signal and $\bsym \psi^k \in \mathbb{H}^n$ the vectorization of the $k$-the wavelet. The HPR then entails solving the optimization problem 
\begin{equation}
    \minimize_{\widetilde{\mathbf{x}}\in\mathbb{H}^n} \frac{1}{2nL}\sum_{k=1}^{L} \left(\left\vert\bsym{\psi}^k\star\widetilde{\mathbf{x}} \right\vert^2 - \left\vert\bsym{\psi}^k\star\mathbf{f} \right\vert^2\right)^2.
\end{equation}
The selection of the quaternion mother wavelet may reduce the ill-posedness of this QWT PR. For instance, quaternion Haar functions have shown promising applications in multi-resolution image analysis. Similarly, quaternion Gabor wavelets arise in applications such as optical flow estimation. Some common choices of quaternion mother wavelets are built upon combining the QFT and conventional complex-valued wavelets such as Haar, biorthogonal, Daubechies, and Coiflet. Similar problems are found in octonion wavelet PR that employs octonion WT. 

\subsection{Sedenion Phase Retrieval (SPR)}
Consider $x \in \mathbb{S}$ denote an unknown sedenion-valued signal, expressed as
$$
x=x_0+\sum_{i=1}^{15} x_i e_i,
$$
where $x_0 \in \mathbb{R}$ and $x_k \in \mathbb{R}$ are the scalar and vector components, respectively. Consider a set of known sedenion-valued sensing vectors $\left\{a_m\right\}_{m=1}^M \subset \mathbb{S}$. The sedenion phase retrieval problem consists of recovering $x$ (up to inherent ambiguities) from intensity-only measurements
$$
y_m=\left|\left\langle a_m, x\right\rangle\right|^2=\left\langle a_m, x\right\rangle\left\langle a_m, x\right\rangle^*, \quad m=1, \ldots, M,
$$
where $(\cdot)^*$ denotes sedenion conjugation and $\langle\cdot, \cdot\rangle$ is a suitably defined sedenion inner product. Due to the non-commutativity, non-associativity, and non-alternativity in the sedenion algebra, the SPR problem is ill-posed and requires development of suitable sedenion calculus and additional structure or constraints.

\section{Summary}
This chapter presented an emerging framework for HPR aimed at recovering high-dimensional signals arising in modern optical imaging and computational sensing applications. By extending conventional complex-valued phase retrieval formulations to hypercomplex domains, HPR leverages the underlying Clifford algebraic structure to jointly process correlated signal dimensions. This holistic treatment not only improves reconstruction performance in many settings but also enables new theoretical insights and recovery guarantees that are unavailable in traditional complex-valued PR methods.

Building on classical PR formulations, we discussed several hypercomplex analogues corresponding to commonly used measurement models. In particular, a broad class of HPR problems can be formulated using structured hypercomplex sensing operators, such as those derived from linear canonical transforms (LCTs), with explicit formulations provided for specific definitions of the QFT/OFT. Extending these formulations to alternative QFT and OFT definitions remains an open and important research direction. We further highlighted scenarios in which PR requires higher-order statistics, such as bispectrum analysis, especially when the underlying signals are nonlinear or non-Gaussian—as encountered in radar, microscopy, and holography. In this context, bispectrum-based HPR naturally arises in applications such as color object recognition.

Recently, fractional Fourier transform (FrFT)-based PR has gained traction in optical imaging and radar systems. While FrFT-based PR introduces additional rotational ambiguities compared to conventional Fourier-based formulations, hypercomplex extensions, such as quaternion FrFT-based PR, offer a promising avenue for addressing these challenges in multidimensional settings.

Beyond measurement design, we discussed the role of structure-promoting priors in HPR, including low-rank and sparse representations, which are well motivated by hypercomplex models of RGB image denoising and multispectral image reconstruction. In parallel, recent advances in data-driven phase retrieval open new opportunities for learning-based HPR using quaternion and octonion neural networks. Finally, we identified the co-design of diffractive optical elements and hypercomplex reconstruction algorithms as a key open problem, with the potential to significantly enhance reconstruction accuracy in future optical sensing systems.

Overall, hypercomplex phase retrieval represents a nascent yet rapidly evolving research area at the intersection of signal processing, applied algebra, and imaging science. As sensing systems continue to grow in dimensionality and complexity, HPR provides a principled and flexible framework for structure-aware phase retrieval beyond the limits of conventional approaches.

\section*{Acknowledgements}
The authors are sincerely grateful to Roman Jacome for his contributions to developing HPR problem via the article \cite{jacome2024invitation}.